\documentclass[letterpaper, 10 pt, conference]{ieeeconf}
\IEEEoverridecommandlockouts 

\usepackage{times}
\usepackage{multicol}
\usepackage[bookmarks=true]{hyperref}

\usepackage{amsmath,amssymb,amsfonts}
\usepackage{graphicx}
\usepackage{textcomp}
\usepackage{xcolor}
\usepackage{eurosym}
\usepackage{array}
\usepackage{subfigure}
\usepackage{booktabs}
\usepackage{multirow}

\usepackage{algorithm}
\usepackage[noend]{algpseudocode}

\algdef{SE}[DOWHILE]{Do}{doWhile}{\algorithmicdo}[1]{\algorithmicwhile\ #1}%

\usepackage{float}
\usepackage{psfrag}

\definecolor{gray1}{gray}{0.7}
\definecolor{gray2}{gray}{0.9}

\begin{document}

\title{\LARGE \bf Decentralized Dynamic Task Allocation in Swarm Robotic Systems for Disaster Response \\{\large\normalfont EXTENDED ABSTRACT}}

\author{Payam Ghassemi$^{1}$, David DePauw$^{2}$, and Souma Chowdhury$^{3}$
\thanks{$^{1}$Ph.D. Student, Dept. of Mechanical and Aerospace Engineering, University at Buffalo, Buffalo, NY 14260, USA.
        {\tt\small payamgha@buffalo.edu}}%
\thanks{$^{2}$BS Student, Dept. of Mechanical and Aerospace Engineering, University at Buffalo, Buffalo, NY 14260, USA.
        {\tt\small daviddep@buffalo.edu}}%
\thanks{$^{3}$Assistant Professor, Dept. of Mechanical and Aerospace Engineering, University at Buffalo, Buffalo, NY 14260, USA. \textit{Corresponding Author}. 
        {\tt\small soumacho@buffalo.edu}}%
}

\maketitle

\begin{abstract}
Multiple robotic systems, working together, can provide important solutions to different real-world applications (e.g., disaster response), among which task allocation problems feature prominently. Very few existing decentralized multi-robotic task allocation (MRTA) methods simultaneously offer the following capabilities: consideration of task deadlines, consideration of robot range and task completion capacity limitations, and allowing asynchronous decision-making under dynamic task spaces. To provision these capabilities, this paper presents a computationally efficient algorithm that involves novel construction and matching of bipartite graphs. Its performance is tested on a multi-UAV flood response application.
\end{abstract}
\begin{keywords}
Multi-robotic task allocation, unmanned aerial vehicles, flood response.
\end{keywords}
\IEEEpeerreviewmaketitle
\section{INTRODUCTION}\label{sec:intro}
Coordinating tasks among collaborative multi-robot systems that must operate without conflict calls for efficient \textit{multi-robot task allocation} (MRTA) methods~\cite{gerkey2004formal,ghassemi2018decentralized,ghassemi2019decentralized}. 
While centralized approaches to solving MRTA problems have traditionally dominated the fields of robotics, transportation, and IoT~\cite{korsah2013comprehensive, colistra2014problem}, decentralized methods have gained increasing prominence in recent years. This is partly due to concerns regarding the scalability of purely centralized approaches and their vulnerability to communication disruptions~\cite{wu2011online}, and partly driven by accelerated advancements in robot autonomy capabilities~\cite{siciliano2016springer}. In this paper, we develop a novel computationally-efficient decentralized algorithm that not only tackles the above challenges but also demonstrates applicability to scenarios with asynchronous decision-making and dynamic tasks (i.e., new tasks appear during mission). This problem falls into the Multi-task Robots, Single-robot Tasks, and Time-extended Assignment (MR-ST-TA) class defined in \cite{gerkey2004formal}.

The performance of the new approaches are compared with that of a centralized ILP based approach and biased random-walk baseline.
The next section presents our proposed decentralized MRTA framework. 
Results, encapsulating the performance of these methods on different-sized problems and a parametric analysis of the proposed decentralized method, are presented in Section~\ref{sec:results}. The paper ends with concluding remarks.
\section{DECENTRALIZED MRTA ALGORITHM} \label{sec:probDecent}
Algorithm~\ref{alg:decMRTA} depicts the pseudocode of our proposed decentralized MRTA or \textbf{Dec-MRTA} algorithm. Each robot is assumed to run the Dec-MRTA algorithm at each decision-making step (e.g., 1 min before finishing its current task) to take the best action that maximizes the team's (mission) outcome. 
Our Dec-MRTA algorithm is composed of three components, which are described next. 
\noindent\textbf{1) Weighted Bipartite Graph Construction:} In order to represent and analyze the task-robot relations, we use the concept of bipartite graphs, or bigraphs. 
A bigraph is a graph whose vertices can be divided into two sets such that no two vertices in the same set are joined by an edge~\cite{asratian1998bipartite}. In this paper, we define our problem as a weighted bigraph $(\mathsf{R, T, E})$ during each decision time-period, where $\mathsf{R}$ and $\mathsf{T}$ are a set of robots and a set of tasks, respectively; and $\mathsf{E}$ represents a set of weighted edges that connect robots to available tasks. 

\noindent\textbf{2) Bigraph Weights Assignment - Robots' Incentive Model:} In order to fully construct the representative weighted bipartite graph, we should determine the weights of edges, a typically challenging endeavor given the lack of any standard recommendations to this end. In other words, the purpose of weighted bigraph construction is to identify and systematically represent the incentive of robots for doing each task, in a manner that facilitates mission success. In this paper, the mission outcome (goal) is defined as delivering the survival kits to the maximum possible number of victims prior to their respective time deadlines. 

We handcraft the incentive (graph edge weight) model to be a negative exponential function of the time ($t_i^r$) by which the robot $r$ can accomplish the concerned task $i$ if chosen next, and if and only if the task can be completed before the deadline $\delta_i$, i.e., if $t_i^r\le\delta_i$; this function is scaled by a remaining flight-range parameter ($\Delta_r$). If the task cannot be completed before the deadline, a weight of zero is assigned. With this model, the weight, $w_{ri}$, of a bigraph edge $(r,i)$ can be expressed as:
\vspace{-0.2cm}
\begin{equation}
\label{eq:edgeWeight}
w_{ri} = 
  \begin{cases}
    \max{(0,\Delta_r - \epsilon)}\cdot \exp{\left(-\frac{t_{i}^r}{\alpha}\right)} & \quad \text{if } t_{i}^r \leq \delta_i\\
    0  & \quad \text{Otherwise}
  \end{cases}
\vspace{-0.2cm}
\end{equation}
%
where $\Delta_r = l^r - (d_{ri}+d_{i0})$. Here, $l^r$, $d_{ri}$, and $d_{i0}$ respectively represent the remaining range of the UAV $r$ at that time instant, the distance to be traveled by robot $r$ to get to task $i$, and the distance between task $i$ and the depot. The parameter $\alpha$ is a normalizing constant (scaling length) for time and the margin parameter $\epsilon$ is the lowest remaining range that a UAV is allowed to travel with. The scaling factor $\Delta_r$ is designed to regulate the remaining range (to undertake further tasks) after the completion of the selected task $i$. The robots are assumed to all start from/end at a single depot. At the beginning, the robots' labels are randomly assigned. 
 


\noindent\textbf{3) Maximum Weight Matching:} Once the weighted bipartite graph has been constructed, the final step 
is to solve the task assignment or allocation problem as a maximum weight matching problem~\cite{west2001introduction}. This problem is defined as drawing a largest possible set of edges such that they do not share any vertices and the summation of the weights of the selected edges are maximum. An improved maximum matching algorithm~\cite{galil1986efficient} is used here to determine the optimal task assignment. 
It is important to note that the outcomes of this (uniquely) asynchronous decentralized decision-making process are deterministic and inherently conflict free.
%
{
\begin{algorithm}[!t]
\caption{Dec-MRTA Algorithm}\label{alg:decMRTA}
\small \textbf{Input:} $\mathcal{T}^k,\mathcal{S}^k$ - the recent states of active tasks and the robots, including robot-$r$ ($\mathcal{S}_r^k$) and its peers ($\mathcal{S}_{-r}^k$).\\
\textbf{Output:} $\mathcal{A}_r^k$ - the next decision of robot-$r$ at its iteration $k$.
\begin{algorithmic}[1]
\State {$\mathbf{A}_r^k \gets 0$} \Comment{return to the depot}
\If{robot-payload $> 0$ \textbf{and} $\Delta_r\geq\epsilon$} 
\State {$\mathcal{T}_r^{k+1} \gets $ \textsc{getFeasibleTask}($\mathcal{T}^k,\mathcal{S}_{r}^k$)}
\If {$\mathcal{T}_r^{k+1} \neq \emptyset$}
\For{$1 \leq i \leq m, \; i \neq r$}
\State {$\mathcal{T}_i^{k+1} \gets $ \textsc{getFeasibleTask}($\mathcal{T}^k,\mathcal{S}_{i}^k$)}
\EndFor
\State {$\hat{\mathcal{T}}^{k+1} \gets \cup_{i=1}^{m}\mathcal{T}_i^{k+1}$}
\State {$\mathbf{G} \gets $ \textsc{constructGraph}($\hat{\mathcal{T}}^{k+1}, \mathcal{S}^k$)}
\State {$\mathcal{A} \gets $ \textsc{maxMatchGraph}($\mathbf{G}$)}
\State {$\mathbf{A}_r^k \gets \mathcal{A}[r]$} \Comment{$\mathcal{A}$ shows decisions of all robots}
\EndIf
\EndIf
\State {\textbf{return} {$\mathbf{A}_r^k$}}
\Procedure{getFeasibleTask}{$\mathcal{T},\mathcal{S}_r$}
\State {$\mathcal{T}_\text{feasible} \gets \emptyset$}
\For{$i \in \mathcal{T}$}
\State {$t_i^r \gets $ global time that robot-$r$ finishes task-$i$}
\State {$\Delta_r \gets $ avail. range of robot-$r$ after doing task-$i$}
\State {$w_{ri} \gets $ Using robots' incentive model, Eq.\eqref{eq:edgeWeight}}
\If{$t_i^r \leq \delta_i$ \textbf{and} $ \epsilon \leq \Delta_r$}
\State {$\mathcal{T}_\text{feasible} \gets \mathcal{T}_\text{feasible} \cup \{\mathcal{T}_i, t_i^r, w_{ri}\}$}
\EndIf
\EndFor
\State {\textbf{return} {$\mathcal{T}_\text{feasible}$}}
\EndProcedure
\end{algorithmic}
\end{algorithm}
}
\section{RESULTS AND DISCUSSION}\label{sec:results}
In this paper, the delivery of survival kits for flood victims via a UAV team is considered as the application. 
We design and execute a set of numerical experiments to investigate the performance and scalability of the \textit{Dec-MRTA} 
approach, and compare it with a \textit{Feasibility-preserving Random-walk MRTA} (RND-Feas) approach where each robot randomly chooses available and feasible tasks (through a
random allocation that uses \texttt{GetFeasibleTask} in Alg.~\ref{alg:decMRTA}). Moreover, in order to measure the optimality of  decision-making of the Dec-MRTA, a centralized \textit{ILP} 
is run and compared. 

\begin{figure}[!t]
\centering
\includegraphics[width=0.42\textwidth]{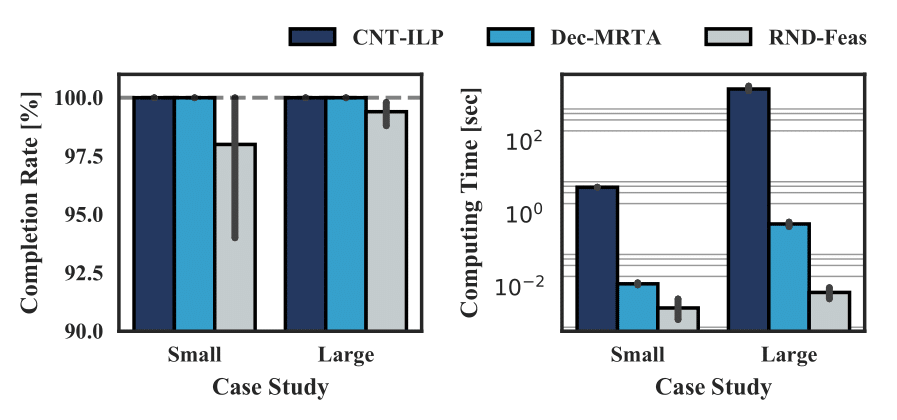}
\vspace{-0.6cm}
\caption{The performance of the algorithms for the two static case studies. \textbf{A log-scale used to show the computing time}. 
The computing time reported for Dec-MRTA and RND-Feas is the commutative computing time.}
\label{fig:CompareMethods}
\vspace{-0.4cm}
\end{figure}

\noindent\textbf{Comparative Analysis of Dec-MRTA:} As shown in Fig.~\ref{fig:CompareMethods}, the completion rate of the centralized ILP and Dec-MRTA algorithms is found to be 100\% in all scenarios, 
while that of the biased random-walk approach is found to vary from 94\% to 100\% across the case scenarios. 
In terms of the computational efficiency, the biased random-walk approach is the fastest. 
\textit{More importantly, as observed from Fig.~\ref{fig:CompareMethods}, the cumulative computing time of Dec-MRTA 
is about 3 orders of magnitude smaller than that of the centralized ILP approach.} 

\begin{figure}[!t]
\centering
\includegraphics[width=0.35\textwidth]{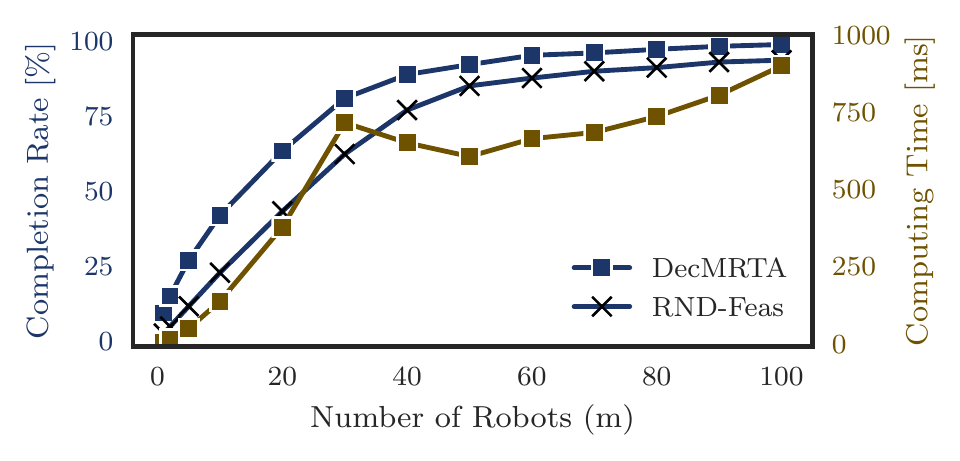}
\vspace{-0.5cm}
\caption{Static case study 1000 tasks: The scalability analysis of the decentralized Dec-MRTA approach. The computing time reported as the average of robots' computing time.}
\label{fig:scalabilityAnalysis}
\vspace{-0.8cm}
\end{figure}
%
\noindent\textbf{Scalability  Analysis of Dec-MRTA:} In order to study the impact of the number of robots (scalability) on computational tractability of the Dec-MRTA algorithm, we tested it for the huge problem and the dynamic case studies by changing the number robots from 1 to 100.
For both case studies, the proposed algorithm outperforms the biased random-walk method in terms of completion rate (the huge problem is shown in Fig.~\ref{fig:scalabilityAnalysis}). The mission success (completion rate) improves by increasing the number of robots, but saturates after certain point (after $m=80$ and $m=50$ in static huge case and dynamic case, respectively).
\noindent\textbf{Communication Latency Analysis of Dec-MRTA:} Here, we run Dec-MRTA on the huge Case with swarm sizes varying from 1 to 100 to elaborate how a 1-minute communication latency impacts on the performance (completion rate) of the Dec-MRTA approach. 
There is no significant impact for 1 and 2-robot swarm case and for swarm with size larger than 50 robots. For 20-robot and 30-robot swarms, the 1-minute latency has a big impact (about $45\%$).
\section{CONCLUSION} \label{sec:conclusion}
In this paper, we proposed a decentralized graph (construction and matching) based algorithm to perform task allocation in multi-robot systems, and assess its performance on a multi-UAV flood response application. 
The new algorithm, Dec-MRTA, is compared with a feasibility-preserving random-walk and a centralized ILP method. Dec-MRTA outperforms the random-walk approach by achieving (up to) 5\% and 57\% better completion rate in a large 100-robot/1000-fixed-task case and a dynamic-task case, respectively. Compared to the ILP method, Dec-MRTA is observed to offer up to $10^3$ times higher computational efficiency, and similar robustness across missions. 
Future work will focus on alleviating the deterministic environment and perfect communication assumptions made in applying Dec-MRTA.
\bibliographystyle{IEEEtran}
\bibliography{payam2018map,asme2ejs}

\appendix       

\section{Centralized ILP MRTA Formulation}

\subsection{Centralized ILP MRTA Formulation} \label{ssec:probCent}
\noindent The centralized MRTA problem is formulated as a Integer Linear Programming (ILP) problem, as given in Eqs.~\eqref{eq:maxVictimMILP} to \eqref{eq:milp_timeDeadline}. It is to be noted that here UAVs are allowed to make multiple tours, and the planning process must not only satisfy the range and payload quantity constraints of each UAV, but also strictly meet the deadline of each task. \textit{To the best of our knowledge, there does not exist such a comprehensive centralized MILP/ILP formulation of the MRTA problem, which can handle multi-robot/multi-tour planning while meeting all given constraints: robot physical constraints (limited range and limited payload) and task deadline constraint.}\\
The decision-space of the ILP comprises two types of binary decision variables, $x_{ijs}^r$ and $y_{is}^r$, where $x_{ijs}^r \in \{0,1\}$, with $x_{iis}^r, x_{i0s}^r, x_{(n+1)0s}^r=0,$ and $y_{is}^r\in\{0,1\}$. The variable $y_{is}^r$ becomes 1 if robot $r$ at the sequence $s$ takes task $i$, and becomes $0$ otherwise. The variable $x_{ijs}^r$ is 1 if robot $r$ takes task $j$ right after finishing task $i$. 
Each robot has a limited payload capacity $Q$ (i.e., maximum tasks per tour) and a limited range $\Delta_\text{range}$; $\delta_{i}$ and $t_{ij}$ represent the time deadline of task $i$ and the time required to finish task $j$ after performing task $i$; $d_{ij}$ is the cost metric for taking task $j$ after performing task $i$. 
\begin{align}
\label{eq:maxVictimMILP}
\max_{x_{ijs}^r,y_{is}^r} \sum_{s\in \mathbf{H}}\frac{1}{s}\sum_{r \in \mathcal{R}} \sum_{i \in \hat{\mathcal{T}}} y_{is}^r
\end{align}
subject to
\begin{align}
\label{eq:milp_noDoubleVisit}
&\sum_{j\in \hat{\mathcal{T}}^e} x_{ijs}^r = y_{is}^r; \quad i\in \mathcal{T}, s \in \mathbf{H}, r \in \mathcal{R}\\
\label{eq:milp_flowBalance}
&\sum_{i\in \mathcal{T}} x_{iks}^r - \sum_{j\in \hat{\mathcal{T}}^e} x_{kjs}^r = 0; \quad k\in \hat{\mathcal{T}}, s \in \mathbf{H}, r \in \mathcal{R}\\
\label{eq:milp_routeStart}
&\sum_{j\in \hat{\mathcal{T}}^e} x_{0js}^r = 1; \quad s\in\mathbf{H}, r\in\mathcal{R}\\
\label{eq:milp_noSubtour}
&\sum_{i,j\in {\mathcal{T}}^e} x_{ijs}^r \leq \sum_{i\in \hat{\mathcal{T}}} y_{is}^r + 1; \quad s\in\mathbf{H}, r\in\mathcal{R}\\
\label{eq:milp_robotPerTask}  
&\sum_{r \in \mathcal{R}}\sum_{s \in \mathbf{H}}  y_{is}^r \leq 1; \quad i\in \hat{\mathcal{T}}\\
\label{eq:milp_arcPerRobotSeq} 
&\sum_{r \in \mathcal{R}}\sum_{s \in \mathbf{H}} x_{ijs}^r \leq 1; \quad i, j\in \hat{\mathcal{T}}\\
\label{eq:milp_limitedCapacity}
&\sum_{i\in \hat{\mathcal{T}}} y_{is}^r \leq \mathbf{Q}; \quad s\in \mathbf{H}, r \in \mathcal{R}\\
\label{eq:milp_limitedRange}
&\sum_{i,j\in \mathcal{T}^e} d_{ij}x_{ijs}^r \leq \Delta_\text{range}; \quad s\in \mathbf{H}, r \in \mathcal{R}\\
\label{eq:milp_timeDeadline}
&\sum_{i,j\in \mathcal{T}^e}\sum_{s\in\{1..s'\}} t_{ij}x_{ijs}^{r} \geq \delta_{i'} y_{i'(s'+1)}^{r}; i' \in \hat{\mathcal{T}}, s'\in \hat{\mathbf{H}}, r \in \mathcal{R}
\end{align}
\noindent Here $\mathcal{R} = \{1..m\}$ is a finite non-empty set of robots, and $\mathcal{T} = \{0..n\}$ is a finite non-empty set of active tasks that each robot is allowed to take, including the depot (index $0$). In Eqs.~\eqref{eq:maxVictimMILP} to \eqref{eq:milp_timeDeadline}, $\hat{\mathcal{T}}=\mathcal{T}-\{0\}, \mathcal{T}^e=\{n+1\}\cup\mathcal{T}, \mathcal{H}=\{1..h\}, \hat{\mathcal{H}}=\mathcal{H}-\{h\}$, and $h$ represents the maximum number of tours each robot is allowed to undertake. 

\end{document}